# Phase Transition in Size- and Charge-Asymmetric Model Electrolytes

A. L. Khomkin[*] and I. A. Mulenko[**]

*IVTAN (Institute of High Temperatures) Scientific Association,
Russian Academy of Sciences, Moscow, 125412 Russia
**Ukrainian State Marine Technical University, Nikolaev, Ukraine

Abstract
A theoretical model of vapor–liquid phase transition in a system of charged hard cores of different diameters is suggested (with the parameters of the transition obtained in a number of studies using the Monte Carlo method). The model is based on the assumption that, in the neighborhood of the critical point, the system of charged cores is a mixture of multipolarly interacting neutral complexes.

## INTRODUCTION

Recent publications [1–4] describe the use of the numerical Monte Carlo method and modern techniques of parallel computation to investigate the parameters of vapor–liquid phase transition in systems representing a mixture of positively and negatively charged hard cores. Mixtures of cores of different diameters were treated with $\sigma_+$ and $\sigma_-$, respectively. The calculations were performed for both charge-symmetric $(Z_+/Z_- = 1:1)$ and charge-asymmetric (2:1 and 3:1) [3] model electrolytes. Yan and Pablo [4] further examined the effect of the asphericity of charges on the parameters of phase transition. It was observed that the dependences of the critical parameters on the asymmetry factor $\lambda = \sigma_+/\sigma_-$ contradict the traditional Debye mean-field theory which predicts the increase of the critical parameters with decreasing $\lambda$ [1–5].

In [1–4], the classical system of hard cores with the particle interaction potential is treated,

$$U_{i,j}(r) = \begin{cases} +\infty & r_{i,j} < \sigma_{i,j}, \\ \pm Z_+ Z_- q^2 / r_{i,j} & r_{i,j} > \sigma_{i,j}, \end{cases} \qquad \sigma_{i,j} = (\sigma_i + \sigma_j)/2. \qquad (1)$$

The obtained values of the critical temperature of transition turn out to be very low compared to the characteristic energy of interaction of unlike charge upon their close approach $u_0 = Z_+ Z_- q^2 / \sigma_{+-}$. The critical temperature in a system of equal and charge-symmetric cores is $0.049 u_0$; for cores of different sizes, this temperature decreases. If one uses the analogy with the hydrogen plasma and assume that $u_0$ corresponds to the ionization potential, the critical temperature turns out to be less than 8000 K. Under these conditions, both in the hydrogen plasma and in the system of cores being treated, all charges will be found as a minimum in two-particle bound states. It is this assumption that underlies the suggested interpretation of the results of [1–4].

## BASIC EQUATIONS.

We will first treat the charge-symmetric (1:1) case and assume that a system of charged cores at close-to-critical temperatures is a gas of Bjerrum doublets in the "ground" state, i.e., when the charges come close together. In this case, a pair of unlike charges is a dipole with the dipole moment $\mu = q\sigma_{+-}$.

The interaction between doublets is described by the potential $V(r,\vartheta_1,\vartheta_2,\varphi_2-\varphi_1)$ [6],

$$V(r,\vartheta_1,\vartheta_2,\varphi_2-\varphi_1) = \infty, \qquad r,\vartheta_1,\vartheta_2,\varphi_2-\varphi_1 \subset \Omega$$
$$V(r,\vartheta_1,\vartheta_2,\varphi_2-\varphi_1) = -\frac{\mu^2}{r^3}g(\vartheta_1,\vartheta_2,\varphi_2-\varphi), \qquad r,\vartheta_1,\vartheta_2,\varphi_2-\varphi_1 \subset \overline{\Omega} \qquad (2)$$

$$g(\vartheta_1,\vartheta_2,\varphi_2-\varphi_1) = 2\cos\vartheta_1\cos\vartheta_2 - \sin\vartheta_1\sin\vartheta_2\cos(\varphi_2-\varphi_1) \qquad (3)$$

In Eqs. (2) and (3), $r, \vartheta_1, \varphi_1, \vartheta_2, \varphi_2$ denote the distance between the centers of dipoles and the angular coordinates of their orientation, respectively. The function given by Eq. (3) describes the dependence of the interaction potential on the dipole orientation. The region $\Omega$ in Eq. (1) corresponds to the values of spherical and angular coordinates in the case of which a close contact of the doublets being treated occurs. Accordingly, $\overline{\Omega}$ denotes the entire volume, less $\Omega$.

We will construct the Van der Waals equation for a gas consisting of Bjerrum doublets. For this purpose, we will use the simplest qualitative procedure described by Landau and Lifshitz [7] and based on extrapolating the expression for the second virial coefficient. The second virial coefficient $B(T)$, corresponding to the doublet interaction has the form [6]

$$B(T) = -\frac{1}{4}\int_0^\infty\int_0^{2\pi}\int_0^\pi\int_0^\pi (\exp(-V(r,\theta_1,\vartheta_2,\varphi_2-\varphi_1)/kT)-1)\sin\vartheta_1 d\vartheta_1 \sin\vartheta_2 d\vartheta_2 d(\varphi_2-\varphi_1)r^2 dr, \qquad (4)$$

where $k-$ is the Boltzmann constant, and $T-$ is the temperature. We will follow Landau and Lifshitz [7] and divide $B(T)$ into two parts $B_1(T)$ and $B_2(T)$ which correspond the repulsion and attraction of doublets,

$$B_1(T) = -\frac{1}{4}\int_\Omega (-1)\sin\vartheta_1 d\vartheta_1 \sin\vartheta_2 d\vartheta_2 d(\varphi_2-\varphi_1) r^2 dr, \qquad (5)$$

$$B_2(T) = -\frac{1}{8}\int_{\overline{\Omega}} \left(\frac{\mu^2}{r^3 kT}g(\vartheta_1,\vartheta_2,\varphi_2-\varphi)\right)^2 \sin\vartheta_1 d\vartheta_1 \sin\vartheta_2 d\vartheta_2 d(\varphi_2-\varphi_1) r^2 dr. \qquad (6)$$

In writing $B_2(T)$ in the expansion of the Boltzmann component in Eq. (4), Eq. (6) retains the first term which gives a nonzero value. It must be emphasized that, in contrast to the conventional procedure [7], when $B_2(T)$ is proportional to the particle interaction potential, in the case of dipole–dipole interaction, $B_2(T)$ turns out to be proportional to the square of that potential, because the angle-averaged value of the linear term is zero. The second virial coefficient for a hard-core system with a dipole moment was calculated by Hirschfelder al. [6]; however, we cannot directly use this result, because the dipoles being treated are soft dumbbells.

In order to avoid the very complex integration in Eqs. (5) and (6) with respect to the angular and radial variables, we will estimate $B_1(T)$ using the fact that this quantity by its physical meaning is some excluded volume defined by the integral of the square of radius. It may be quite safely assumed that $B_1(T)$ will be proportional to the total volume of the doublet,

$$B_1(T) = \frac{2}{3}\pi k_1 (\sigma_+^3 + \sigma_-^3) = \frac{2}{3}\pi \sigma_1^3. \tag{7}$$

In calculating $B_2(T)$, the square of interaction potential, proportional to the sixth power of $1/r$, is averaged over distances, and it is necessary to reasonably estimate the effective distance of the least approach of doublet charges. We will assume that this distance is proportional to the doublet diameter,

$$\sigma_2 = k_2(\sigma_+ + \sigma_-). \tag{8}$$

As a result, we have, for $B_2(T)$,

$$B_2(T) = -\frac{2}{9}\pi \frac{\mu^4}{\sigma_2^3 (kT)^2}. \tag{9}$$

We will write the equation of state for a system consisting of $N$ doublets located in volume $V$ at temperature $T$ and pressure $P$,

$$P = \frac{NkT}{V}\left(1 + \frac{N}{V} B(T)\right). \tag{10}$$

By dividing the second virial coefficient into repulsion and attraction parts and introducing an excluded volume, one can change from Eq. (10) to the Van der Waals equation,

$$P = \frac{NkT}{V - Nb_1} - \left(\frac{N}{V}\right)^2 b_2 \left(\frac{\varepsilon_0^2}{kT}\right)\frac{1}{3}. \tag{11}$$

Conventional notation convenient for further uses is introduced into Eq. (11),

$$b_1 = \frac{2}{3}\pi\sigma_1^3, \tag{12}$$

$$b_2 = \frac{2}{3}\pi\sigma_2^3, \tag{13}$$

$$\varepsilon_0 = \frac{\mu^2}{\sigma_2^3}. \tag{14}$$

One can use the conventional procedure [7] and, based on Eq. (11), derive expressions for the critical density $\rho_{cr}$ and temperature $T_{cr}$ of vapor–liquid phase transition,

$$\rho_{cr} = \frac{N_{cr}}{V} = \frac{1}{3b_1}, \tag{15}$$

$$kT_{cr} = \frac{2\sqrt{2}}{9}\varepsilon_0\sqrt{\frac{b_2}{b_1}}. \tag{16}$$

For the dimensionless density $\rho^* = 2\rho\sigma_{+-}^3$ and temperature $T^* = T\sigma_{+-}/q^2$, we derive

$$\rho_{cr}^* = \frac{1}{2\pi k_1}f(\lambda), \tag{17}$$

$$T_{cr}^* = \frac{\sqrt{2}}{18\sqrt{k_1}k_2^{3/2}}\sqrt{f(\lambda)}, \tag{18}$$

where

$$f(\lambda) = \frac{1}{4}\frac{(1+\lambda)^3}{(1+\lambda^3)}. \tag{19}$$

We will determine the coefficients $k_{1,2}$ in Eqs. (7) and (8) using the data on the parameters of phase transition for cores of equal sizes ($\lambda = 1$) $\rho_{cr}^* = 0.073$, $T_{cr}^* = 0.0492$. As a result, we obtain $k_1 = 2.18$; $k_2 = 1.054$. The obtained value of the coefficient $k_2$ indicates that the distance of the least approach of two doublets is selected quite reasonably and corresponds to the distance of the least approach of two oppositely charged cores, which also corresponds to the distance of the least approach of two Bjerrum doublets. The coefficient $k_1$ fully coincides with the Ishihara factor $f$ [6] introduced in the calculation of the second virial coefficient for solid aspherical particles $B_m(T) = 4v_m f$,, where $v_m$ is the particle volume. For spherical particles, this factor is unity; for a cube, 11/8; and for a regular tetrahedron, 1.926. If we calculate the second virial coefficient for hard cores of a diameter equal to the total diameter of the doublet $\sigma_+ + \sigma_-$, the Ishihara factor will be four. These estimates lead one to assume that a Bjerrum doublet in the neighborhood of the critical point behaves rather as a randomly rotating dumbbell which takes up a volume close to that of a tetrahedron.

Equations (17) and (18) yield

$$\rho_{cr}^* = 0.073 f(\lambda), \tag{20}$$

$$T_{cr}^* = 0.0492 \sqrt{f(\lambda)}. \tag{21}$$

CALCULATION RESULTS

Curves corresponding to relations (20) and (21) are given in Fig. 1 along with the data of numerical experiment of Yan and Pablo [2]. The latter data are qualitatively and quantitatively correctly described by these relations. Note that the critical temperature is less dependent in the asymmetry factor than the critical density. This is associated with the root dependence of (18) on the asymmetry function $f(\lambda)$ and is due to the fact that the attraction component of pressure (11) is inversely proportional to temperature; in the traditional Van der Waals equation, this component is independent of temperature.

We will now treat a system of charge-asymmetric cores [3, 4] of different sizes. It is natural to assume that this system of cores in the vicinity of the critical point (because the critical transition temperatures are low compared to $u_0$ will consist of simplest neutral complexes, namely, triplets for the (2:1) option and quadruples for (3:1). A triplet which has the least binding energy is a complex consisting of a positively charged core with two adherent negative cores spaced at the maximum possible distance $\sigma_+ + \sigma_-$ from each other. A quadruple consists of a positively charged core with three adherent negative cores likewise spaced at the maximum possible distance from each other, i.e., their centers are spaced on a circumference at 120 degrees from one another. The total charge and dipole moment of such complexes are zero; they exhibit cylindrical symmetry. The interaction between these complexes is quadrupole-quadrupole. The potential of such interaction decreases over long distances as $1/r^5$ and contains, as a factor, a fairly complex function of angular variables, the average of which is zero. It is only the result of averaging the square of this function that is other than zero. We will derive the critical parameters of phase transitions in these systems using similarity relations to fit the option of charge-symmetric cores examined above. The quadrupole moments $Q$ of the complexes are $4q\sigma_{+-}^2$ for the case of (2:1) and $1.5q\sigma_{+-}^2$ for (3:1). The dependence of the constant $\varepsilon_0$ on the magnitude of the charge and the half-sum of diameters will not change ($\varepsilon_0 \approx Q^2/\sigma_{+-}^5 \approx q^2/\sigma_{+-}$), and neither will relation (8) for $\sigma_2$; naturally, Eq. (7) for $\sigma_1$ must change because of the variation of the volume taken up by the complex,

$$\sigma_1 = k_1(Z_+\sigma_-^3 + \sigma_+^3). \tag{22}$$

One can readily write expressions for the critical parameters,

$$\begin{aligned}\rho_{cr}^*(Z_+,\lambda) &= \rho_{cr}^*(Z_+,1)f(Z_+,\lambda),\\ T_{cr}^*(Z_+,\lambda) &= T_{cr}^*(Z_+,1)\sqrt{f(Z_+,\lambda)},\\ f(Z_+,\lambda) &= \frac{Z_+ + 1}{8}\frac{(1+\lambda)^3}{(Z_+ + \lambda^3)}.\end{aligned} \tag{23}$$

These relations are normalized such that, in the case of cores of equal diameters ($\lambda=1$), they agree with the values obtained in numerical experiments for $\rho_{cr}^*(Z_+,1)$ and $T_{cr}^*(Z_+,1)$. They are valid for charge-symmetric cores ($Z_+ = 1$) as well. Curves corresponding to relations (23) are given in Figs. 2 and 3 along with the data of numerical experiment. The variable in these figures is provided by the quantity $\delta = (\lambda - 1)/(\lambda + 1)$, corresponding to that introduced by Panagiotopoulos and Fisher [3]. Relations (23) demonstrate qualitative agreement with the results of numerical experiment. In addition to reasonable dependences on the asymmetry parameter, they correctly describe the shift of the maximum of dependence of the critical parameter on the factor $\delta$, which was recorded in the numerical experiment. The positions of these maxima adequately correspond to the maxima of $f(Z_+,\delta)$ as functions of $\delta$. The solution of the equation $\partial f(Z_+,\delta)/\partial\delta = 0$ gives $\delta_m(Z_+ = 1) = 0$; $\delta_m(Z_+ = 2) = 0.172$; $\delta_m(Z_+ = 3) = 0.268$.

The obtained value of the coefficient $\kappa_1 = 2.18$ indicates that, for further development of the model, one must take into account the effect of the rotation of complexes on their effective size. The rotation of complexes will undoubtedly bring about some effective increase in their size, as follows from the results of [8] as well.

## CONCLUSIONS

Expressions for the parameters of vapor–liquid phase transition recorded in numerical experiments have been suggested on the basis of the assumption about the composition of a low-temperature system of charged cores. The suggested theoretical interpretation of the results of numerical experiments is based on the assumption that the system of charged cores at low temperatures is a mixture of multipolarly interacting neutral complexes. The obtained dependence of the parameters of phase transition on the asymmetry parameter qualitatively and, in a number of cases, quantitatively agrees with the results of numerical simulation by the Monte Carlo method.

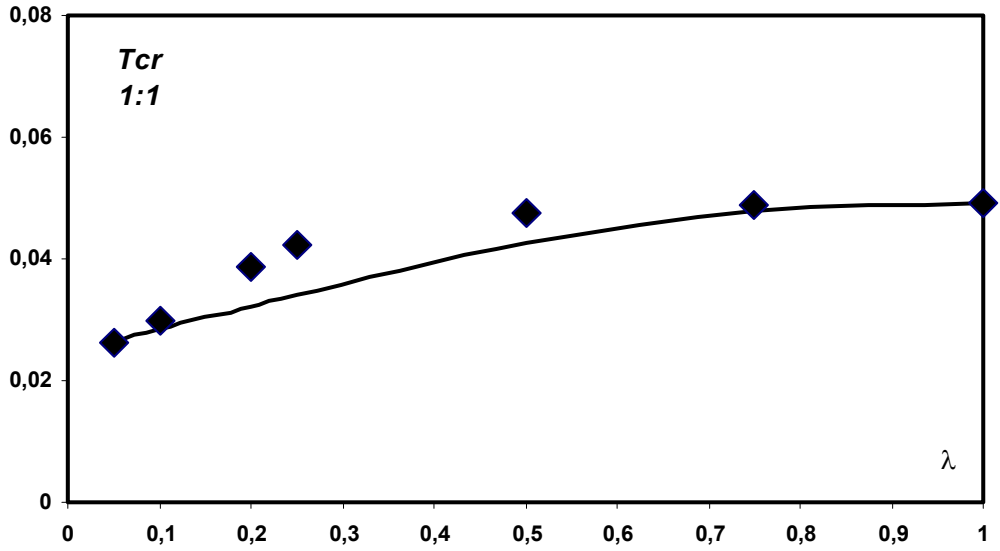

а)

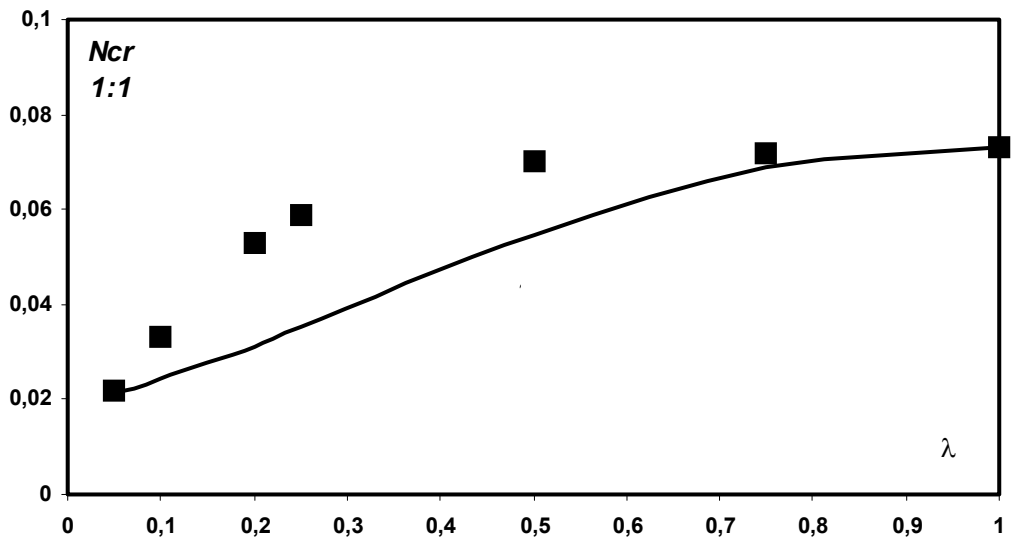

б)

Fig. 1. The dependence of (a) reduced critical temperature and (b) density on the asymmetry factor for charge-symmetric model electrolytes: points, numerical experiment of [2]; curve, our results.

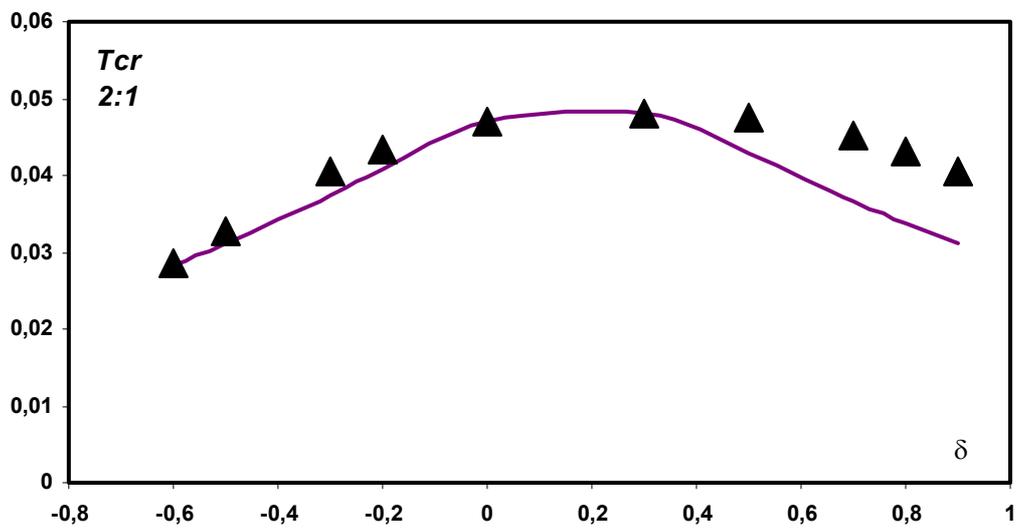

а)

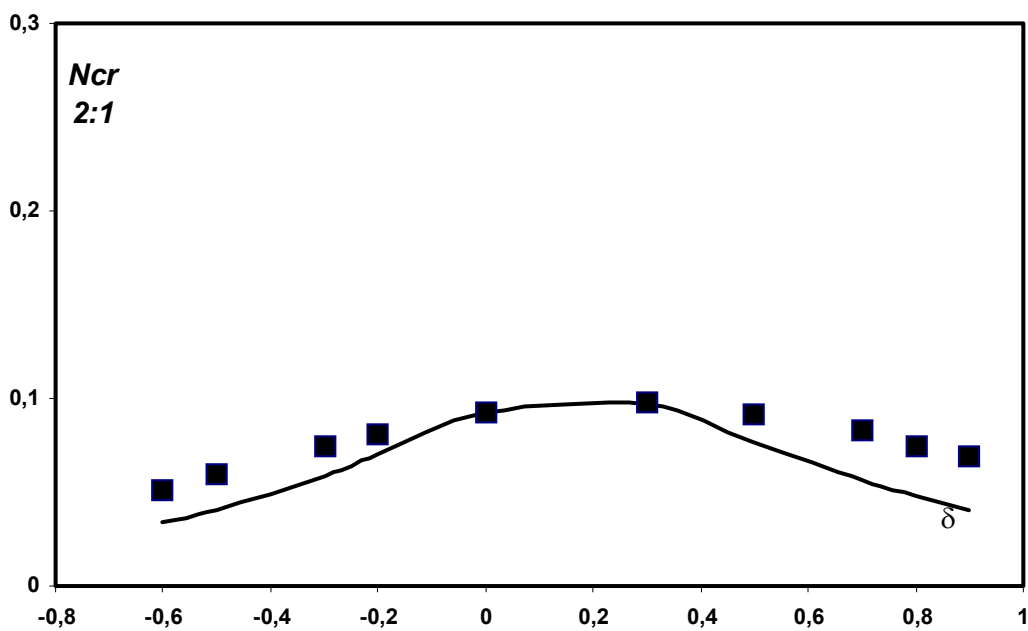

б)

Fig. 2. The reduced (a) critical temperature and (b) density for a 2:1 model: points, numerical experiment of [3]; curve, our results.

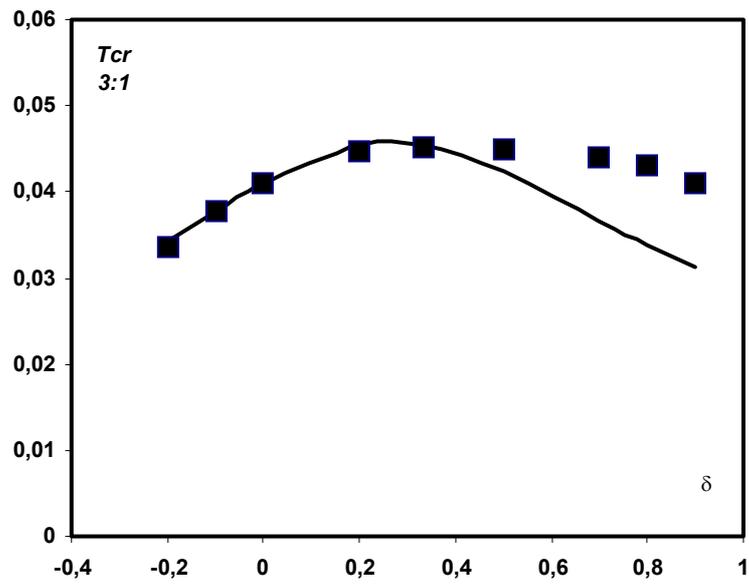

а)

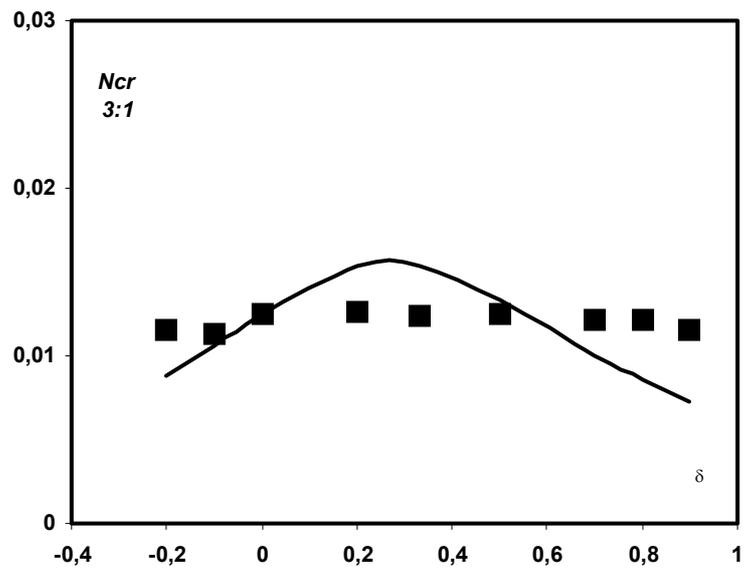

б)

Fig. 3. The reduced (a) critical temperature and (b) density for a 3:1 model. Designations are the same as in Fig. 2.